\documentclass[letter]{aa} 
\usepackage{graphicx}
\usepackage{txfonts}
\usepackage{color}
\usepackage{natbib}
\usepackage{nameref}

\newcommand{\Msun}{M$_{\odot}$}

\newcommand{\gppr}{\stackrel{>}{\scriptstyle \sim}}
\newcommand{\gappr}{\raisebox{-0.4ex}{$\gppr$}}

\newcommand{\mesa}{MESA}

\newcommand{\cv}{CV}
\newcommand{\cvs}{CVs}

\newcommand{\msun}{M$_\odot$}

\defcitealias{garraffoetal18-1}{CG18a}
\defcitealias{garraffoetal18-2}{CG18b}

\begin{document} 

   \title{Suggested magnetic braking prescription derived from field complexity fails to reproduce the cataclysmic variable orbital period gap}
\titlerunning{Suggested MB based on field complexity does not reproduce CV period gap}

   \author{Valentina Ort\'uzar-Garz\'on
          \inst{1}
          \and
          Matthias R. Schreiber 
          \inst{1}
          \and
          Diogo Belloni
          \inst{1}
          }

   \institute{Departamento de F\'isica, Universidad T\'ecnica Federico Santa Mar\'ia, Avenida Espa\~na 1680, Valpara\'iso, Chile
             }

   \date{Received \ Accepted }

  \abstract
   {Magnetic wind braking drives the spin-down of low-mass stars and the evolution of most interacting binary stars. A magnetic braking prescription that was claimed to reproduce both the period distribution of
   cataclysmic variables (CVs) and the evolution of the rotation rates of low-mass stars is based on a relation between the angular momentum loss rate and magnetic field complexity. } 
   {The magnetic braking model based on field complexity has been claimed to predict a detached phase that could explain the observed period gap in the period distribution of \cvs~but has never been tested in detailed models of \cv~evolution. Here we fill this gap. }
   {We incorporated the suggested magnetic braking law in MESA and simulated the evolution of \cvs~for different initial stellar masses and initial orbital periods.} 
   {We find that the prescription for magnetic braking based on field complexity fails to reproduce observations of \cvs. The predicted secondary star radii are smaller than measured, and an extended detached phase that is required to explain the observed period gap (a dearth of non-magnetic \cvs~with periods between{ ${\sim}2$ and ${\sim}3$ hours}) is not predicted. 
   }
   {Proposed magnetic braking prescriptions based on a relation between the angular momentum loss rate and field complexity are too weak to reproduce the bloating of donor stars in \cvs~derived from observations and, in contrast to previous claims, do not provide an explanation for the observed period gap. The suggested steep decrease in the angular momentum loss rate does not lead to detachment. Stronger magnetic braking prescriptions and a discontinuity at the fully convective boundary are needed to explain the evolution of close binary stars that contain compact objects. The tension between braking laws derived from the spin-down of single stars and those required to explain \cvs~and other close binaries containing compact objects remains.}


   \maketitle
%

\section{Introduction}

Magnetic fields of low-mass stars force the mass lost in winds to co-rotate with 
the star up to the Alfv\'en radius. This causes the terminal specific angular momentum of the wind to be higher 
than the specific angular momentum of the stellar surface. 
The resulting angular momentum loss is called magnetic wind braking and represents a fundamental ingredient 
of stellar astrophysics. 
Magnetic braking drives the spin-down of single stars \citep{schatzman62-1,mestel68-1} 
and the secular evolution of virtually all close binary stars \citep[see][for a recent review]{belloni+schreiber23-1}. 
{Despite this importance, magnetic braking remains poorly understood. In particular, different prescriptions are used for single stars and stars in binaries.}

Decades ago, \citet{skumanich72-1} found that the rotational periods ($P_{\mathrm{rot}}$) of Sun-like stars are proportional to the square root of their age, which translates to an angular momentum loss of $\dot{J}\propto\,M\,R^4P_{\mathrm{rot}}^{-3}$ (with $M$ and $R$ representing the mass and the radius of the star).  
However, observations of chromospheric activity, coronal X-ray emission, flare activity, and magnetic field strengths in low-mass main-sequence stars reveal that these observables are all correlated and increase with rotation only up to a mass-dependent critical rotation rate. For shorter periods, the relation between activity and rotation saturates 
\citep[e.g.][]{staufferetal94-1,reinersetal09-1,medinaetal20-1}.
The assumption that these observables also relate with magnetic braking led to saturated magnetic braking prescriptions being postulated, in which the dependence of the magnetic braking torque on the spin period becomes shallower above a given rotation rate \citep[e.g.][]{sillsetal00-1,andronovetal03-1}.

More recent measurements of low-mass main-sequence stars in young clusters revealed a bimodal distribution of fast and slower rotation rates that is difficult to explain with Skumanich-like or saturated magnetic braking prescriptions 
\citep[e.g.][]{meibometal11-1,newtonetal16-1}. 
A relation between the strength of magnetic braking 
and the field complexity has been suggested by \citet{garraffoetal16-1} 
to potentially solve this issue \citep[][hereafter, \citetalias{garraffoetal18-1}]{garraffoetal18-1}. 

In close binary stars with orbital periods shorter than ${\sim}10$\,days, tidal forces cause the stellar rotation to be synchronised with the orbit {\citep[e.g.][]{levato74-1,meibometal06-1,flemingetal19-1}}. A Skumanich-like magnetic braking prescription therefore predicts very strong orbital angular momentum loss in these close binaries 
\citep{rappaportetal83-1}. 
These high angular momentum loss rates for close binaries represent a key ingredient for the standard evolution theory of cataclysmic variables (CVs). According to this scenario, \cvs~with donor stars that still contain a radiative core experience strong angular momentum loss due to magnetic braking, which causes the donor stars to be bloated and the mass transfer rates to be high. When the secondary star becomes fully convective, at an orbital period of ${\sim}3$ hours, magnetic braking becomes much weaker, which allows the donor star to shrink and detach from its Roche lobe. The binary evolves as a detached system towards shorter periods until the mass transfer rate resumes at a much lower rate at a period of ${\sim}2$\,hours. 

In other words, to produce a detached phase that covers a period range {from ${\sim}2$ to ${\sim}3$\,hours,} two conditions need to be fulfilled. First, above the gap, the donor star needs to be significantly oversized compared to its equilibrium radius. Second, at the fully convective boundary, the mass transfer timescale needs to become longer than the radius adjustment timescale of the donor star. 
In the standard scenario for \cv~evolution, the first condition requires strong magnetic braking above the gap, while the second condition is met by assuming a drastic decrease in magnetic braking at the fully convective boundary
\citep[see e.g.][for more details]{belloni+schreiber23-1}.

Recently, \citet{schreiberetal21-1} showed that the late appearance of white dwarf magnetic fields \citep{bagnulo+landstreet21-1}, possibly related to a crystallisation-driven dynamo \citep{isernetal17-1,schreiberetal21-2,schreiberetal22-1,ginzburgetal22-1,schreiberetal23-1,bellonietal24-1}, affects the evolution of \cvs~and should cause a reduction in magnetic braking if the field is strong enough. The evolution described above might thus fully apply only to \cvs~with weakly or non-magnetic white dwarfs. For \cvs~containing strongly magnetic white dwarfs, so-called polars, the white dwarf magnetic field can reduce the wind zones of the secondary star, thereby significantly reducing magnetic braking. 
This reduction has been predicted to cause a much less pronounced period gap, or none at all \citep[see also][]{webbink+wickramasinghe02-1,bellonietal20-1}. 

The evolution of \cvs~is observationally constrained by the measured orbital period distribution \citep{kniggeetal11-1,inightetal23-1,inightetal23-2,schreiberetal24-1} and the mass transfer rate, which can be determined from the mass-radius relation of the donor stars \citep{kniggeetal11-1,mcallisteretal19-1} or the temperature of the white dwarf \citep{palaetal17-1,palaetal22-1}. The orbital period distribution of \cvs~containing weakly or non-magnetic white dwarfs indeed shows a dearth of systems with periods between 
$\sim2$ and $\sim3$ hours. This period gap seems to be at least much less pronounced (maybe even absent) for \cvs~containing strongly magnetic white dwarfs \citep{schreiberetal24-1}.
The mass transfer rates of \cvs~measured from donor star radii are significantly higher at periods longer than three hours \citep{kniggeetal11-1,mcallisteretal19-1}. These observations are roughly consistent with the mass transfer rates determined from white dwarf temperatures \citep{palaetal22-1}. 

These observational constraints 
support the idea that the field of a strongly magnetic white dwarf reduces magnetic braking. For non-magnetic CVs, the observations agree reasonably well with predictions made assuming a Skumanich-like 
magnetic braking for donor stars that still contain a radiative core, and significantly weaker magnetic braking for fully convective stars \citep{kniggeetal11-1,bellonietal18-1}. 

However, \cv~evolution has turned out to be difficult to explain when applying more recent prescriptions derived for single stars. 
In an attempt to unify magnetic braking prescription for single stars and close binaries, 
\citet[][hereafter, \citetalias{garraffoetal18-2}]{garraffoetal18-2} applied their magnetic braking prescription based on field complexity to the evolution of \cvs~and claimed that it is possible to reproduce the observed orbital period distribution. If true, this result would greatly reduce the tension between magnetic braking prescriptions derived for single stars and those used in binary evolution studies. 

Here we use detailed binary calculations to show that the 
magnetic braking prescription suggested by \citetalias{garraffoetal18-2} can explain neither the existence of the period gap for \cvs~containing weakly or non-magnetic white dwarfs nor the large radii derived from observations of such \cvs~above the gap. We then discuss possible avenues towards a unified magnetic braking prescription.

\section{Binary star simulations}

We used the \mesa~code \citep{Paxton2011, Paxton2013, Paxton2015, Paxton2018, Paxton2019, Jermyn2023}, {version r-23.03.1,} to compute the evolution of CVs\footnote{Inlists available at \href{https://zenodo.org/doi/10.5281/zenodo.13632684}{Zenodo}}.
The \mesa~ equation of state is a blend of the OPAL \citep{Rogers2002}, SCVH
\citep{Saumon1995}, FreeEOS \citep{Irwin2004}, HELM \citep{Timmes2000},
PC \citep{Potekhin2010}, and Skye \citep{Jermyn2021} equations of state.
Radiative opacities are primarily from OPAL \citep{Iglesias1993,
Iglesias1996}, with low-temperature data from \citet{Ferguson2005}
and the high-temperature, Compton-scattering dominated regime from
\citet{poutanen17-1}. Electron conduction opacities are from
\citet{Cassisi2007} and \citet{blouinetal20-1}.
Nuclear reaction rates are from JINA REACLIB \citep{Cyburt2010}, NACRE \citep{Angulo1999}, and
additional tabulated weak reaction rates \citep{Fuller1985, Oda1994,
Langanke2000}. Screening is included via the prescription of \citet{Chugunov2007}.
Thermal neutrino loss rates are from \citet{Itoh1996}.
Roche lobe radii in binary systems are computed using the fit of
\citet{Eggleton1983}. Mass transfer rates in Roche lobe
overflowing binary systems are determined following the
prescription of \citet{Ritter1988}.

In our simulation we furthermore assumed the white dwarf mass to be constant, that is, that the same amount of mass that is accreted during a nova cycle is expelled during the eruption, in rough agreement with model predictions \citep{yaronetal05-1}.
The mass expelled during nova eruptions was assumed to carry the specific angular momentum of the white dwarf. While 
this likely underestimates the true value for \cvs~containing lower-mass white dwarfs \citep{schreiberetal16-1}, using this form of consequential angular momentum loss should not
lead to significantly different predictions for
the secular evolution of most \cvs, which contain relatively massive white dwarfs ($\gappr0.8$\Msun).
Our simulations take into account angular momentum loss through gravitational radiation according to \citet{Paczynski1967}. The orbital angular momentum loss through magnetic braking 
suggested by \citetalias{garraffoetal18-1}
is summarised in what follows.

\section{Magnetic braking and field complexity relations revisited}

The magnetic braking prescription we tested is based on a relation between angular momentum loss and the complexity of the magnetic field. This relation has been claimed to explain not only the spin-down rates of single stars \citepalias{garraffoetal18-1} but also the \cv~orbital period distribution \citepalias{garraffoetal18-2}. 
This prescription for angular momentum loss  through magnetic braking $(\dot{J}_{\mathrm{MB}})$, 
developed initially by \citet{garraffoetal16-1}, 
can be written as follows: 
\begin{equation}\label{eq:j}
    \dot{J}_{\mathrm{MB}} = \dot{J}_{\mathrm{Dip}} Q_{\mathrm{J}}(n), 
\end{equation}\label{eq:om}
where 
\begin{equation} 
Q_{\mathrm{J}}(n)=4.05\,e^{-1.4n} + (n -1)/(60 B n),
\end{equation}\label{eq:Q}%
with $B$ representing the field strength. 
The dipole angular momentum loss rate is given by
\begin{equation}
    \dot{J}_{\mathrm{Dip}}=\Omega^3\,c\,\tau,
\end{equation}
where $\Omega = 2\pi/P_\mathrm{orb}$ is the angular velocity, and $c$ is a constant related to the wind efficiency  and is assumed to be $1 \times 10^{41}$\,g\,cm$^2$ \citepalias{garraffoetal18-1}. Finally, the field complexity 
parameter, $n$, is defined as
\begin{equation}\label{eq:n}
    n = a\frac{\tau}{P_\mathrm{rot}} + b\frac{P_\mathrm{rot}}{\tau} + 1. 
\end{equation}
{The parameters $a$ and $b$ were set to $0.02$ and $2$ in \citetalias{garraffoetal18-1}, respectively. However, when applying their model to \cvs,~slightly different values ($a=0.01$\ and $b=1$) were adopted \citepalias[][their Sect. 2]{garraffoetal18-2} with the justification 
that the parameter values were not well constrained in their previous study, which 
only dealt with stars with masses greater than $0.3$\,\Msun. As the aim of this Letter is to test the magnetic braking prescription that was suggested to reproduce the orbital period distribution of \cvs, we used $a=0.01$ and $b=1$.}

For close binary stars, the rotational period, $P_{\mathrm{rot}}$, is equal to the orbital period, $P_{\mathrm{orb}}$ (i.e. the orbit  is synchronised). The dependence of the convective turn-over time on the stellar mass is given by the empirical relation of \citet{wrightetal11-1}: 
\begin{equation}
\log(\tau) = 1.16 - 1.49 \log(M/\mathrm{M}_{\odot}) - 0.54 \log^2(M/\mathrm{M}_{\odot}).
\end{equation}
{Admittedly, this frequently used empirical dependence might need to be updated using larger, recently established datasets \citep{jaoetal22-1}. However, given that we want to test the prescription calibrated by \citetalias{garraffoetal18-2} for \cvs, we followed them in adopting the above relation.   
We furthermore note that this empirical convective turn-over time is different from the calculated global convective turn-over time 
used in alternative magnetic braking prescriptions \citep[e.g.][their Eq. 16]{van+ivanova19-1}. }

Applying their magnetic braking algorithm to \cvs, \citetalias{garraffoetal18-2} 
dropped the term that depends on the field strength in Eq.\,\ref{eq:Q} 
as it is negligible as long as $n$ remains below $7,$ which is the case for the parameter space relevant to the orbital period gap in \cvs. Combining the above equations then results in a largely simplified angular momentum loss prescription through magnetic braking:
\begin{equation}
    \dot{J}_{\mathrm{MB}} =- 4.05~e^{-1.4n}~c~\Omega^3 \tau.    
\end{equation}

To simulate \cv~evolution, \citetalias{garraffoetal18-2} then assumed a mass-radius relation for the donor stars in \cvs~derived from observations \citep[][]{kniggeetal11-1}.     
This last step of their procedure is potentially problematic because it swaps cause and effect. In \cvs,~magnetic braking causes angular momentum loss, which drives the mass transfer, and it is this mass loss of the donor star that determines the mass-radius relation of the donor star. In other words, angular momentum loss (largely through magnetic braking) determines the mass-radius relation of CV donors, which therefore cannot be assumed a priori.

\section{CV evolution driven by magnetic braking related to field complexity}

\begin{figure}
    \resizebox{\hsize}{!}{\includegraphics{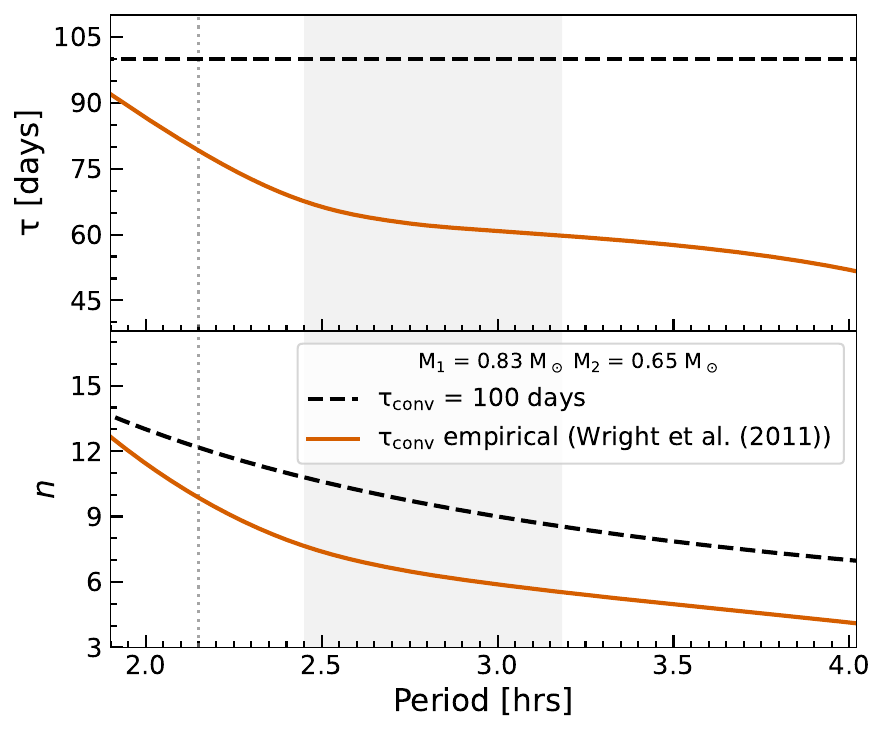}}
    \caption{Comparison of the turn-over time, $\tau,$ evolution (upper panel) based on the empirical relation proposed by \cite[solid orange line]{wrightetal11-1} and a constant value of $100$ days (dashed black line) and the corresponding magnetic field complexities, $n$ (lower panel). The shaded region corresponds to the period gap for non-polar \cvs~according to \cite{schreiberetal24-1}. 
    The dashed grey vertical line indicates the lower period gap edge according to \cite{kniggeetal11-1}.
    The assumption of a constant turn-over time is not justified, and it is incompatible with the condition $n<7$.}
    \label{fig:tau_n}
\end{figure}

We show in Fig.\,\ref{fig:tau_n} the dependence of the field complexity parameter and the convective turn-over time on the orbital period 
assuming the \citetalias{garraffoetal18-2} magnetic braking prescription
for a white dwarf mass of 0.83~\msun\,and an initial donor mass of 0.65~\msun.   
The bottom panel confirms that $n$ indeed remains below ${\sim}7$ beyond the period gap if the formula from \citet{wrightetal11-1} is used for the turn-over time. However, \citetalias{garraffoetal18-2} state that they obtained qualitatively identical results by assuming a constant convective turn-over time of $100$\,days. This is not the case in our simulations. 
The assumption that $n$ remains below $7$, and that the dropped term in 
Eq. \ref{eq:Q} is thus indeed negligible, cannot be justified.

\begin{figure}
    \resizebox{\hsize}{!}{
    \includegraphics[width = \linewidth]{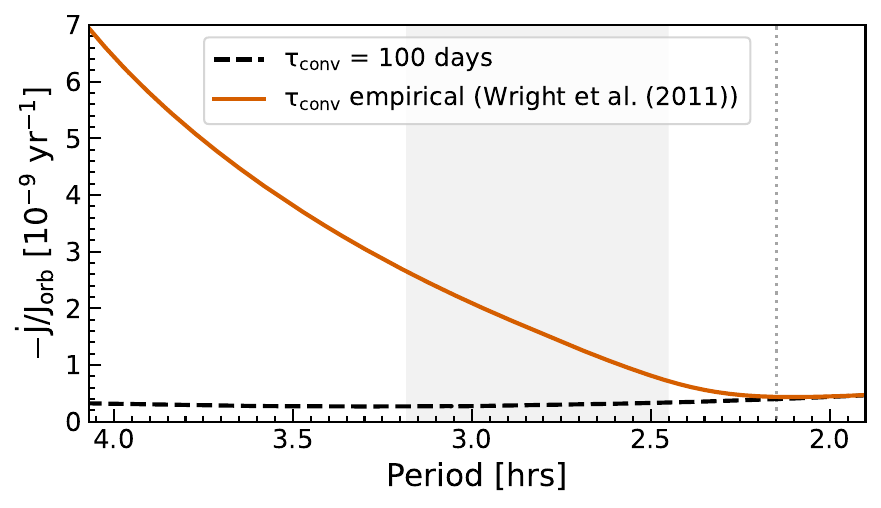}}
    \caption{Relative angular momentum loss of \cvs~with $M_1 = 0.83\,\mathrm{M}_\odot$ and $M_2 = 0.65\,\mathrm{M}_\odot$, with a constant $\tau$ = 100 days (dashed black line) calculated using the empirical relation from \cite[solid orange line]{wrightetal11-1}. 
    The shaded region corresponds to the period gap for non-polar \cvs~according to \cite{schreiberetal24-1}. 
    The dashed grey vertical line indicates the lower period gap edge according to \cite{kniggeetal11-1}.
    }
    \label{fig:jdot_n_values}
\end{figure}

In Fig.\,\ref{fig:jdot_n_values} we show the relative angular momentum loss we obtain using the magnetic braking prescription of \citetalias{garraffoetal18-2}. The strength of angular momentum loss is similar to theirs if the dependence of the convective turn-over time on mass \citep{wrightetal11-1} is taken into account. For a constant value, a convective turn-over time of $100$\,days, the angular momentum loss rate becomes much lower, clearly inadequate to describe \cv~evolution. 

\begin{figure}
    \resizebox{\hsize}{!}{
    \includegraphics{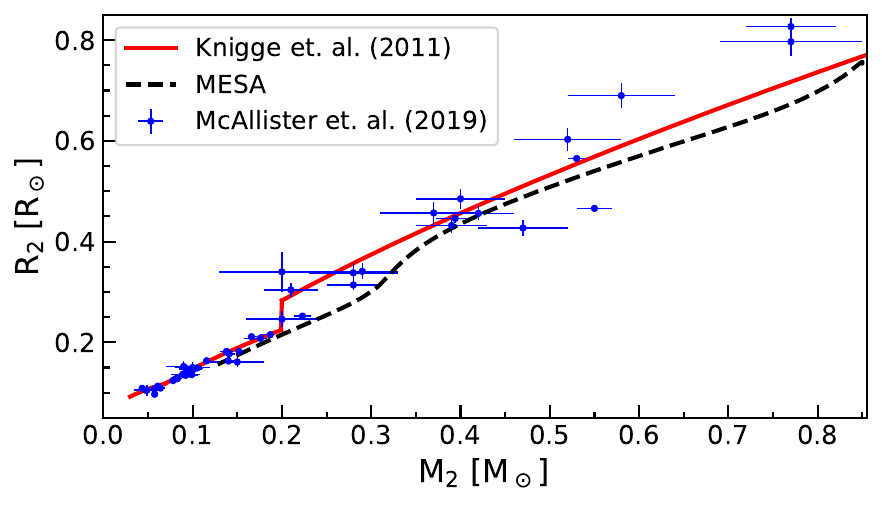}}
    \caption{Donor star radius ($R_2$) as a function of its mass ($M_2$) according to \cite{kniggeetal11-1} (dashed red line) compared to the evolution predicted by MESA for a \cv~with initial $M_1 = 0.9\,\mathrm{M}_\odot$ and $M_2 = 0.85\,\mathrm{M}_\odot$ and assuming magnetic wind braking as in \citetalias{garraffoetal18-2} (solid black line). 
    The radii and masses of \cv~donors derived from observations \citep{mcallisteretal19-1} are shown as blue dots with their respective error bars. It is clear that the prescription based on the complexity of the field fails to reproduce the observations. }
    \label{fig:radius_mass}
\end{figure}

In Fig.\,\ref{fig:radius_mass} we
show the evolution of the donor star radius and mass for a white dwarf mass of 0.9\,\msun, which corresponds roughly to the mean white dwarf mass in \cvs~\citep{zorotovicetal11-1,palaetal22-1}, and an initial donor star mass of 0.85\,\msun, 
assuming the magnetic braking prescription based on field complexity \citepalias{garraffoetal18-2}. 
The initial orbital period was assumed to be $12$ hours. 
Also shown is the mass-radius relation derived by \citet{kniggeetal11-1} from observations of \cvs. 
The predicted radii are significantly smaller than those derived from observations, and the donor star does not sufficiently shrink to detach from its Roche lobe. Instead, the donor star mass decreases continuously.  

\begin{figure}
    \resizebox{\hsize}{!}{
    \includegraphics{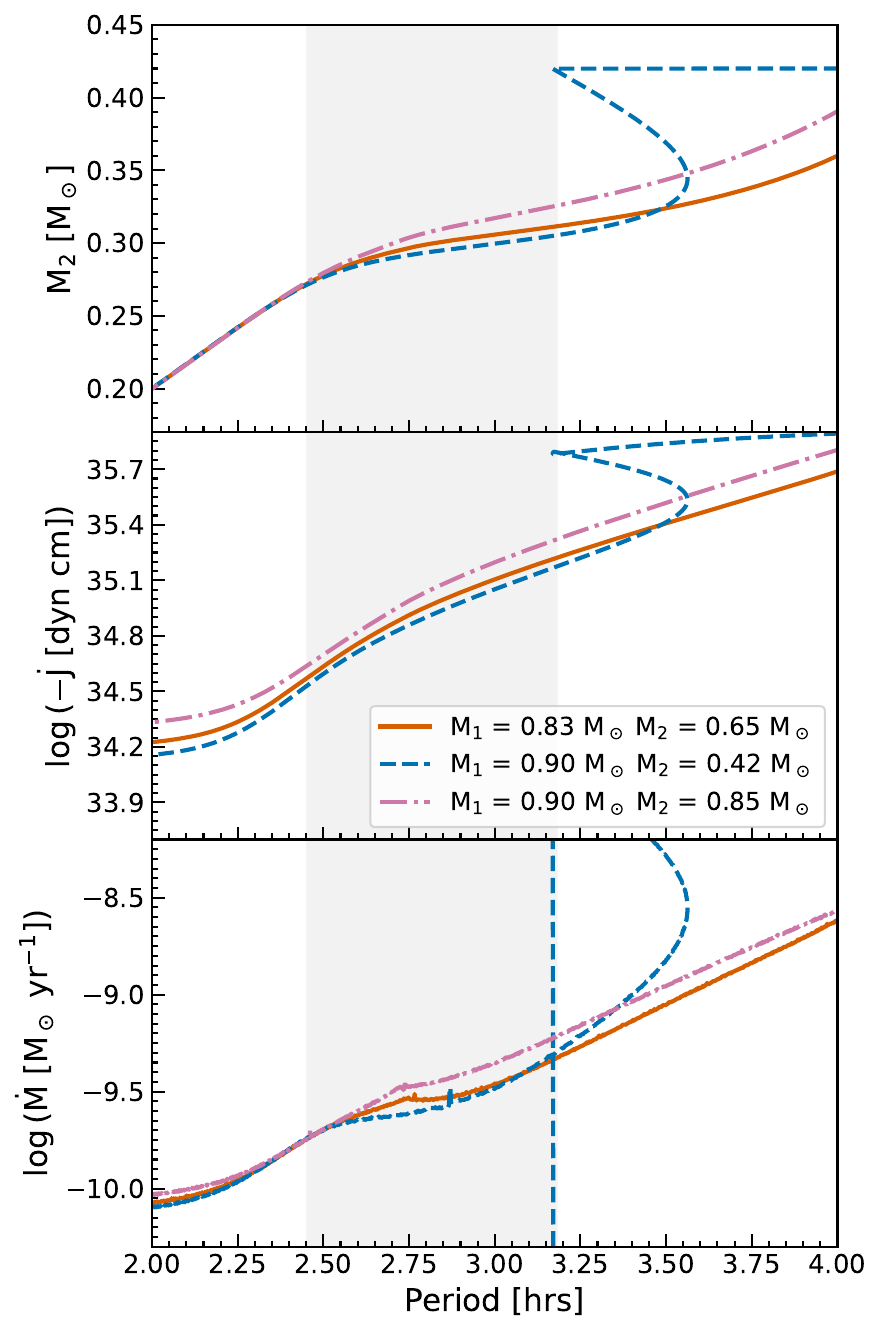}}
    \caption{Evolution of the donor mass (upper panel), angular momentum loss rate (central panel), and mass transfer rate (lower panel) for CVs with $M_{1} = 0.9\,{\rm M}_\odot$, $M_{2} = 0.42\,{\rm M}_\odot$ (dashed blue line),  $M_{1} = 0.83\,{\rm M}_\odot$, $M_{2} = 0.65\,{\rm M}_\odot$ (solid orange line), and $M_{1} = 0.9\,{\rm M}_\odot$, $M_{2} = 0.85\,{\rm M}_\odot$ (dash-dotted pink line). The angular momentum loss rate is calculated as in \citetalias{garraffoetal18-2}.
     The shaded region corresponds to the period gap according to \cite{schreiberetal24-1}. The dashed grey vertical line indicates the lower period gap edge according to \cite{kniggeetal11-1}. 
     {For the lowest-mass donor (blue track), the onset of mass transfer occurs close to the upper edge of the period gap, and the binary evolves through a well-known loop (often called the period flag) at the onset of mass transfer \citep[e.g.][their Fig. 4]{stehleetal96-1}.}  
    As the donor stars do not detach from their Roche lobe, the considered magnetic braking model can not explain this crucial feature in the observed period distribution.}
    \label{fig:mb_garraffo}
\end{figure}

Figure\,\ref{fig:mb_garraffo} shows the evolution of the donor mass, the rate of angular momentum loss due to magnetic braking, and the mass transfer rate as a function of the orbital period for different white dwarf and donor star masses.  
The resulting accretion rates in the considered period range
are roughly of the same order as the observed ones
\citep{palaetal22-1, dubusetal18-1}, but a detached phase that could explain the orbital period gap \citep{kniggeetal11-1, schreiberetal24-1} is not predicted. 
In contrast to the claims by \citetalias{garraffoetal18-2}, our detailed simulations of \cv~evolution show that the steep decrease in angular momentum loss through magnetic braking predicted by their prescription does not cause the systems to detach, and the orbital period gap thus remains unexplained.

\section{Concluding discussion}

Assuming that the efficiency of magnetic braking is related to the magnetic field complexity is very reasonable because the rate of magnetic wind braking must be related to the number of open field lines. 
In particular, given that the field strength saturates in a similar fashion for fully convective stars and those that still contain a radiative core 
\citep{wright+drake16-1}, a dependence of angular momentum loss through magnetic braking on field complexity could represent an elegant way to reproduce the orbital gap.  
This idea goes back several decades \citep{taam+spruit89-1}. 
Finding a prescription for magnetic braking that depends on field complexity and that can reasonably well explain the spin-down of single low-mass stars and key observables of  \cvs~would therefore represent a significant step forwards in our understanding of magnetic braking. This is exactly what has been attempted by 
\citetalias{garraffoetal18-2}, and their results seemed promising. 

However, instead of modelling the detailed evolution of the donor star, 
\citetalias{garraffoetal18-2} assumed a 
mass-radius relation similar to those derived from observations of \cvs~\citep{kniggeetal11-1}, which might not represent the mass-radius relation predicted by their angular momentum loss prescription.  
It is important to note that \citetalias{garraffoetal18-2} were well aware of the limitations of their simulations and stated that more detailed simulations were required.  

Here we filled this gap by implementing their prescription into the stellar evolution code \mesa~and simulating evolutionary tracks of \cvs. 
We find that the magnetic braking model from 
\citetalias{garraffoetal18-2} 
does not allow fundamental observables of \cvs~to be reproduced, such as the donor star radii and the famous orbital period gap. 
The differences between our results and those of \citetalias{garraffoetal18-2} are most likely related to the assumed mass-radius relation. 
Detailed stellar evolution calculations are required to obtain the mass-radius relation for a given prescription of angular momentum loss through magnetic braking since the expansion of the donor star is driven by mass loss, which in turn is driven by angular momentum loss. 
In our detailed simulations, the magnetic braking prescription proposed by 
\citet{garraffoetal15-1} and \citetalias{garraffoetal18-1} does not produce sufficiently bloated donor stars, and the steep decrease in their angular momentum loss rate is not sufficient to cause a detached phase. A more drastic reduction in magnetic braking is required to explain the observed orbital period distribution of \cvs. 

This result emphasises the previously mentioned tension
\citep[e.g.][]{kniggeetal11-1,bellonietal23-1,schreiberetal24-1} between braking laws that describe the rotational evolution of single low-mass stars \citep{mattetal15-1,sillsetal00-1,andronovetal03-1, garraffoetal18-1} and those used to reproduce the evolution of close compact binaries ranging from \cvs~to AM\,CVn stars \citep{belloni+schreiber23-2}, low-mass X-ray binaries \citep{van+ivanova19-1}, or detached white dwarf plus M-dwarf post common envelope binaries \citep{schreiberetal10-1}. 
While relatively weak saturated magnetic braking seems to do relatively well in describing the evolution of single stars, 
much higher angular momentum loss rates are needed to explain the observations of close compact binaries. 
Apart from the different strengths of magnetic braking, a drastic decrease (often assumed to be a discontinuity) at the fully convective boundary is required to explain the \cv~orbital period gap, and such a discontinuity is usually not incorporated into saturated magnetic braking laws. 

Given the recent evidence of saturated magnetic braking in close main-sequence binaries \citep{elbadryetal22-1} and observations of detached white dwarf plus main-sequence binaries \citep{schreiberetal10-1}, which indicate a drastic change at the fully convective boundary similar to that required to explain the \cv~orbital period gap, \citet{bellonietal23-1} developed a disrupted (at the fully convective boundary) and saturated prescription that could simultaneously explain observations of both types of detached binaries. 
A natural next step is to test the very same model for \cvs. 

While this disrupted and saturated magnetic braking prescription represents a promising candidate to describe binary star evolution, one important piece of the puzzle, an explanation for the dramatically different angular momentum loss rates through magnetic braking in single stars and close binaries, is obviously missing. Perhaps the possibility of tidally enhanced mass loss, which is required to explain a number of observed binary star systems \citep{tout+eggleton88-1}, 
should be investigated in more detail.

\begin{acknowledgements}
MRS and DB thank for support from FONDECYT (grant numbers 1221059 and 3220167). VOG and MRS thank C. Garraffo and J. Drake for their insightful feedback. We also thank O. Toloza for her helpful comments and suggestions. 
\end{acknowledgements}

\bibliographystyle{aa} 
\bibliography{bibs}

\end{document}